\documentclass[12pt,preprint]{aastex}

\newcommand{\lsim}{\raisebox{-1ex} {$\; \stackrel{\textstyle <}{\sim}\;$}}

\begin{document}

\title{VORTEX CONFIGURATIONS, OSCILLATIONS AND PINNING IN NEUTRON STAR CRUSTS}

\author{Masaki Hirasawa and Noriaki Shibazaki}
\affil{Department of Physics, Rikkyo University, Nishi-Ikebukuro, Tokyo 171-8501, Japan}
\email{hirasawa@cfr.rikkyo.ne.jp, shibazak@rikkyo.ac.jp}

\begin{abstract}
The rotating neutron superfluid in the inner crust of a neutron star is threaded by quantized vortex lines.  The pinning force from lattice nuclei and the Magnus force from neutron superfluid act onto a vortex line that has a finite tension.  The configuration of a vortex line in equilibrium comprises a number of pinned straight lengths separated by unpinned kinks when the rotation axis is slightly inclined to the major axis of a crystal lattice and the Magnus force is not so strong.  Energy of $\sim 4 \mbox{ MeV}$ is required to form a kink at densities of $\sim 3.4 \times 10^{13}\mbox{ g cm}^{-3}$. Magnus force makes the vortex line parabolic on global scale.  There exist two modes, rotational and helical (Kelvin), for oscillations allowed on a vortex line.  The vortex oscillations in the pinned straight segments are possible only above a minimum frequency.  We find no unstable mode that grows with time.  Hence, the vortex configurations with kinks may be stable and yield the strongly pinned state.  The essence of our results may not be altered even when we consider the vortex behaviors in a polycrystalline structure.  Our studies suggest that pinning of a vortex line in a polycrystalline structure would be still strong enough to explain the large glitches.
\end{abstract}


\section{INTRODUCTION}
Glitches, observed from more than 30 isolated pulsars \citep{lyn00, wan00}, are a sudden jump in pulsar rotation rate.  Slow recovery of a rotation rate following these events implies the presence of loosely-coupled fluids or neutron superfluids in the interior of a neutron star.  The transfer of angular momentum  from the rotating neutron superfluid to the normal matter is likely to be responsible for pulsar glitches and postglitch relaxation.

A rotating neutron superfluid is threaded by quantized vortex lines.  The superfluid can alter its angular velocity by the radial motion of vortices.  In the inner crust of a neutron star neutron rich nuclei that form the bcc lattice coexist with neutron superfluid.  The vortex lines in the inner crust are normally pinned to the lattice nuclei.  Difference in velocity between superfluid and nuclear lattice builds up since magnetic braking slows down the nuclear lattice and pinning prevents the vortex lines from moving.  Then, the neutron star undergoes sudden unpinning of a large number of vortex lines followed by the outward motion.  This catastrophic unpinning has long been considered as a promising cause for glitches \citep{and75, rud76, ala84, pin85, elb92,  bel92}.

A vortex line is subjected to the Magnus force when it moves relative to the superfluid.  The catastrophic unpinning model assumes that the pinning force is strong enough to sustain the vortices to the nuclear lattice against the Magnus force until the moment just before glitches. Based on the condensational and kinetic energies, the pinning energy is estimated, ranging approximately $1-10$ MeV per nucleus \citep{ala84, eps88, piz97}.  This magnitude itself is strong enough against the Magnus force expected just before glitches.  Jones (1992) mentions, however, that the pinning forces on a randomly oriented rigid vortex line largely cancel since there are nearly an equal number of pinning sites on either side of the vortex line.  Link, Epstein, \& Baym (1993) find that for a vortex line of finite tension, pinning becomes much more efficient by slightly bending  and forming kinks. Recently, Jones (1997, 1998, 1999) argues that vortex interaction with a polycrystalline structure does not provide pinning strong enough to explain the large glitches observed in the Vela pulsar (see Section \ref{pinning} for detail).

Here we study the vortex configurations, oscillations and pinning in the inner crust of a neutron star. In Section \ref{equation} we derive the equation of motion of vortex lines.  In Section \ref{configuration} we present the equilibrium configurations of vortex lines.  In Section \ref{oscillation} we examine the stability of equilibrium configurations and the oscillations excited on a vortex line. In Section \ref{pinning} we discuss the vortex pinning based on the results of the previous sections.  In the last section we summarize the results and mention the conclusions.

\section{EQUATIONS OF MOTION OF VORTEX LINES IN AN INNER CRUST}\label{equation}

 We consider the motion of vortex lines in an inner crust, where the lattice nuclei are embedded in a superfluid neutron sea.  We adopt the Cartesian coordinates fixed in the rest frame of the lattice nuclei.  The z-axis is chosen to be parallel to one of the major axes of a body-centered crystal lattice.  In this paper the most attention will be devoted to configurations and oscillations of vortex lines which are parallel to or slightly inclined against the z-axis.  We limit the inclination angle of a vortex line against the z-axis to a small value.  Then, we can describe their motion and configurations by the two-dimensional position vectors, $ {\bf u}=(x, y)$, of vortex lines in the xy plane which depend on the coordinate z.  The equation of motion of vortex lines is written as

\begin{equation}
\textit{\textbf f} _p+{\rho} \mbox{\boldmath$\kappa$} {\times} \left( \frac{{\partial}\textit{\textbf u}} {{\partial}t}-\textit{\textbf v}_s \right)
+T \frac{{\partial}^2\textit{\textbf u}} {{\partial}z^2}=0,\label{eq:eqmot}
\end{equation} 
where $\rho$ is the superfluid density, $\textit{\textbf f}_p$ the pinning force per unit length of a vortex line, $\mbox{\boldmath$\kappa$}$ the circulation of a vortex, $\textit{\textbf v}_s$ the locally uniform superfluid velocity and $T$ the tension of the vortex line.

Vortices in the superfluid are quantized. Each vortex line carries one quantum $\kappa= h/2m$ of circulation, where $h$ is the Planck constant and $m$ the neutron mass. Since we consider the vortex lines aligned or almost aligned with the z-axis, we  approximate the circulation vector by

\begin{equation}
\mbox{\boldmath$\kappa$} = \frac{h}{2m}  \hat{\bf z},
\end{equation}
where $\hat{\bf z}$ is the unit vector along the z-axis.

In equation (\ref{eq:eqmot}) the second term represents the Magnus force which arises when there exists the relative motion between the local superfluid and the vortex.  The magnus force acting on a unit length of a vortex line is given by
 
\begin{equation}
\textit{\textbf f}_M=\rho \mbox{\boldmath$\kappa$} \times (\textit{\textbf v}_v -\textit{\textbf v}_s)
\end{equation}
where $\textit{\textbf v}_v$ is the vortex velocity.  In the vortex unpinning model the
difference in angular velocity between the inner crust superfluid and the
crust is reduced significantly just after the glitches, decoupling the inner
crust superfluid from the crust.  As the crust is slowed down by the
external braking torque, the angular velocity difference $\Delta \Omega$ builds up between the local superfluid and the lattice nuclei.  Using the glitch interval $t_g$ and the angular deceleration rate $\dot{\Omega}$, the angular velocity difference just before the glitch is estimated as

\begin{equation}
\Delta \Omega  \sim |\dot{\Omega}| t_g .
\end{equation}
In the Vela pulsar the angular velocity difference and the Magnus force acting on the pinned vortex line may amount to $\sim 0.01 \mbox{ rad s}^{-1}$ and $ \sim 6.8 \times 10^{14} \mbox{ dyne cm}^{-1}$, respectively, just before the glitches.

The third term in equation (\ref{eq:eqmot}) comes out from the tension of vortices when vortices are distorted from the straight lines.  The energy of vortex lines is equal to the kinetic energy of the velocity field induced by them.  The tension is given by the energy per unit length,

\begin{equation}
T = \rho  \kappa  c_K,
\end{equation}
where the parameter $c_K$ is only logarithmically dependent on the wave number and the neutron superfluid coherence length \citep{son87}.  For numerical estimates we assume $c_K =1.2 \times 10^{-3} \mbox{ cm}^2 \mbox{ s}^{-1}$ \citep{jon98}.

The vortex-nucleus interaction has an attractive component arising
from the change in the condensation energy and a repulsive component arising from the change in kinetic energy of circulating superfluid.  We adopt only the attractive component since we are interested in the inner portion of the inner crust, where the attractive component is dominant over the repulsive component.  The interaction potential we adopt is
\begin{equation}
V = -E_p  \exp \left( -\frac{u^2}{2\xi^2} \right),\label{eq:enpi}
\end{equation}
where $E_p$ is the pinning energy (or the interaction energy) per nucleus, $\xi$ is the vortex core radius or the coherence length of the neutron superfluid and $u$ is the distance in the xy plane between the nucleus and the vortex line, $u^2 = x^2 + y^2$. The coherence length of the neutron superfluid is written as
\begin{equation}
\xi=\frac{2E_F}{\pi k_F \Delta}.
\end{equation}
where $\Delta$ is the energy gap for the superfluid neutrons, $E_F$ their Fermi energy and $k_F$ the Fermi surface wave number \citep{alp77, ala84}.
The pinning force, which acts to pin the vortex line to nuclei, per site is approximately given by
\begin{equation}
F_p=\frac{E_p}{\xi}.
\end{equation}

We can obtain the lattice-vortex interaction potential by summing the above potential (\ref{eq:enpi}) contributed from each nucleus over the entire nuclei of a single large crystal.  Note that the lattice-vortex interaction potential has the translational properties and hence can be expanded into the Fourier series using the reciprocal lattice vectors.  We assume the lattice-vortex interaction potential to be uniform in the z-direction since we are interested in the configurations and oscillations of vortices, which are approximately aligned with the z-axis, with the scale and wave lengths significantly larger than the lattice constant.  Hence, of the twelve basic reciprocal lattice vectors of the bcc lattice we adopt the basic four vectors in the xy plane, $(g,\, g),\, (-g,\, g), \, (-g,\, -g)$ and $(g,\, -g)$, where $g = 2\pi/a$ with $a$ being the lattice constant.  Then, the lattice-vortex interaction potential per unit length of a vortex line is
given by
\begin{equation}
U_p=2 U_g [ \cos(gx+gy)+\cos(-gx+gy)],
\end{equation}
where the Fourier coefficient $U_g$ is $\sim -9.7 \times10^3 \mbox{ erg cm}^{-1}$ at densities of $\rho \sim 3.4\times 10^{13} \mbox{ g cm}^{-3}$.  We obtain the pinning force acting on the unit length of a vortex line by taking the derivative of $U_p$ with respect to $x$ and $y$,
\begin{equation}
f_{px}=2 g U_g [ \sin(gx+gy)-\sin(-gx+gy)] \label{eq:fx}
\end{equation}
and
\begin{equation}
f_{py}=2 g U_g [ \sin(gx+gy)+\sin(-gx+gy)],\label{eq:fy}
\end{equation}
where $f_{px}$ and $f_{py}$ are the $x$ and $y$ components of the pinning force, respectively.

We conduct the coordinate rotation and transform the variables from $(x,\,y)$ to $(\phi,\,\psi)$ defined as

\begin{equation}
\phi=gx+gy \label{eq:phi}
\end{equation}
and
\begin{equation}
\psi=-gx+gy.\label{eq:psi}
\end{equation}
Using equations (\ref{eq:fx}), (\ref{eq:fy}), (\ref{eq:phi}) and (\ref{eq:psi}), the equation of motion of vortices (\ref{eq:eqmot}) reduces to
\begin{equation}
4g^2U_g\sin \phi -\rho \kappa \frac{\partial \psi}{\partial t}+\rho \kappa gv_s+\rho \kappa c_K \frac{\partial ^2 \phi}{\partial z^2}=0 \label{eq:eqmot1}
\end{equation}
and
\begin{equation}
4g^2U_g\sin \psi+\rho \kappa \frac{\partial
 \phi}{\partial t}-\rho \kappa gv_s+
\rho \kappa c_K\frac{ \partial ^2 \psi}{\partial z^2}=0.\label{eq:eqmot2}
\end{equation}
Here and hereafter we assume for simplicity that the neutron superfluid flows in the y-direction. Note that equations (\ref{eq:eqmot1}) and (\ref{eq:eqmot2}) describe the motion of vortices in terms of the components in the directions of reciprocal vectors.

\section{EQUILIBRIUM CONFIGURATIONS OF VORTEX LINES}\label{configuration}

\subsection{Equations for Equilibrium Configurations}

 If the vortex lines are completely rigid, the vortex configuration
would be simply a straight line.  The tension of a superfluid vortex line in
the inner crust of a neutron star, however, is finite.  Hence, when the
vortex lines undergo the forces such as the pinning and Magnus
forces,  the vortex lines in general are bent and their shape is deformed
from the straight line.  We examine the equilibrium configurations of
vortex lines here.

We assume that in equilibrium $\partial \phi / \partial t= 0$ and $\partial \psi / \partial t= 0$ in equations (\ref{eq:eqmot1}) and (\ref{eq:eqmot2})  and hence the vortices are stationary with respect to the nuclear lattice. Then, equations (\ref{eq:eqmot1}) and (\ref{eq:eqmot2}) can be integrated to yield

\begin{equation}
-4g^2U_g \cos{\phi}+{\rho}{\kappa}gv_s{\phi}
+\frac{1}{2}{\rho}{\kappa}c_K\left(\frac{{\partial}
{\phi}}{{\partial}z}\right)^2=\mbox{const}\label{eq:stst-int1}
\end{equation}
and

\begin{equation}
-4g^2U_g \cos{\psi}-{\rho}{\kappa}gv_s{\psi}
+\frac{1}{2}{\rho}{\kappa}c_K\left(\frac{{\partial}
{\psi}}{{\partial}z}\right)^2=\mbox{ const}.\label{eq:stst-int2}
\end{equation}
Once a displacement and orientation of the vortex line are given at , for example, z = 0, we can solve equations (\ref{eq:stst-int1}) and (\ref{eq:stst-int2}) and derive the equilibrium
configurations of vortex lines.

Hereafter, for numerical calculations we use the physical parameters such as the
pinning energy, the coherence length and the lattice constant at densities of $\rho
\sim 3.4 \times 10^{13}\mbox{ g cm}^{-3}$, where a large fraction of crust
superfluids resides and the pinning energy peaks.  We adopt $v_s \sim 10^4 \mbox{
cm s}^{-1}$ for the superfluid velocity unless otherwise stated.

The first, second and third terms in equations (\ref{eq:stst-int1}) and (\ref{eq:stst-int2}) correspond to the pinning force, the Magnus force and the vortex tension, respectively. The pinning force per site is $F_p  \sim   9.6 \times 10^5$ dyne and the vortex tension is $T \sim 8.1 \times 10^7 $ dyne, while the Magnus force acting on the vortex section of a length corresponding to the lattice constant $a$ is $\rho \kappa v_s a \sim 3.9 \times 10^3$ dyne.  We note that the Magnus term is much smaller than the pinning and tension terms.

We should note that equations (\ref{eq:stst-int1}) and (\ref{eq:stst-int2}) are analogous to the energy conservation equation for the particle motion in a potential well. The first and second terms and third term on the left hand side of equations (\ref{eq:stst-int1}) and (\ref{eq:stst-int2}) may correspond to the potential and kinetic energies in the particle motion, respectively, while the integration constant on the right hand side to the total energy.  The gradient of the vortex line against the z-axis may be looked as the velocity of a particle.  The difference between the "total and potential energies" yields the gradient of the vortex line. The pinning term in equation (\ref{eq:stst-int1}) (equation (\ref{eq:stst-int2})) has peaks at $\phi =2n\pi$ ($\psi =2n\pi$) with $n$ being integer, while the Magnus term increases (decreases) linearly with $\phi$ ($\psi$). As easily inferred from this analogy, solutions to equations (\ref{eq:stst-int1}) and (\ref{eq:stst-int2}) can be classified into the three characteristic cases in terms of the magnitude of the integration constant ( the "total energy") relative to the peak values of the potential energy due to the pinning and Magnus forces.

\subsection{Straight Line Solutions}
\begin{figure}
\epsscale{0.70}
\plotone{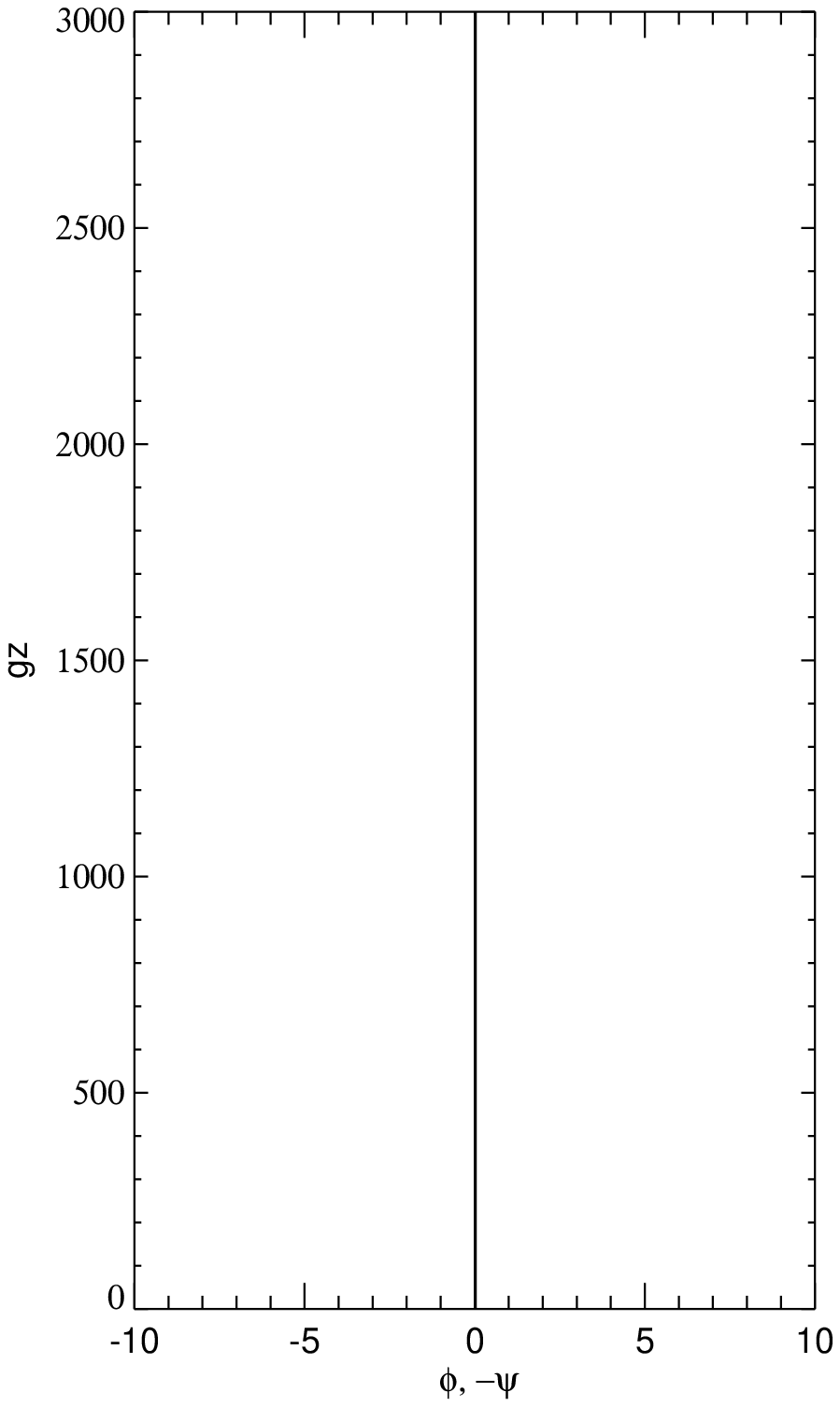}
\caption[f1.eps]{A straight line solution to a vortex equilibrium configuration. The major axis of a single crystal is chosen as a z-axis.  The horizontal coordinate expresses displacement of a vortex segment from the z-axis in the direction of a reciprocal lattice vector.  All coordinates are multiplied by $g = 2\pi /a$.  This straight solution is obtained when the rotation axis of the neutron star is perfectly parallel to the z-axis.  Here the Magnus force is included.  The vortex line passes through almost the bottom of a pinning potential ($\phi = - \psi \sim 0.016$). \label{straight}}
\end{figure}

 Let us consider the case where the integration constant (the "total
energy") in equations (\ref{eq:stst-int1}) and (\ref{eq:stst-int2}) is just equal to a peak value of the sum of pinning and Magnus terms (the "potential energy").  The derivative of $\phi$ and $\psi$ with respect to z should be zero over an entire vortex line. Hence, the equilibrium solution for the vortex configuration is a straight line which is in parallel with the z-axis or the major axis of a single crystal lattice (Fig. \ref{straight}).  Note that the average orientation of a vortex line expresses the direction of the rotation axis.  Thus, the rotation axis of the star is also parallel to the z-axis.  When the Magnus term is neglected in equations (\ref{eq:stst-int1}) and (\ref{eq:stst-int2}), the straight vortex line passes through the bottom of the pinning potential well along the z-axis.  When the Magnus term is included, the straight vortex line, though still parallel to the z-axis, is displaced only slightly from the z-axis, $x \sim 1.4 \times 10^{-14} \mbox{ cm} $ and $y \sim 0 \mbox{ cm} $.  In actual cases when the Magnus force acts on the vortex lines, however, the equilibrium configuration of the vortex may deviate from the straight line, though very slightly, since the pinning sites are located discretely along the z-axis.  We have the strongest pinning regime when the rotation axis happens to be oriented along the major axis of a single crystal since the vortex
line intercepts the pinning sites most often.

\subsection{Sinusoid-like Solutions}
\begin{figure}
\epsscale{0.70}
   \plotone{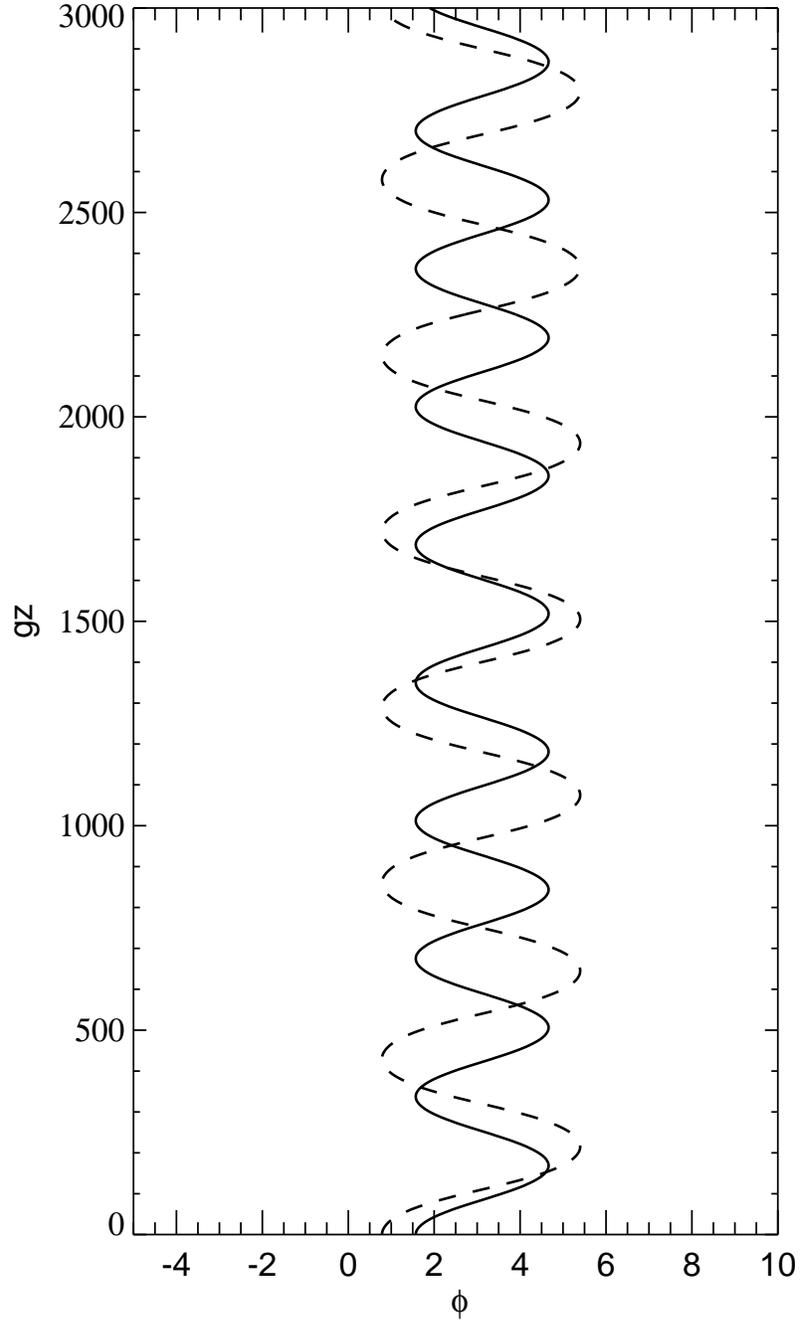}
\caption[f2.eps]{Sinusoid-like solutions to a vortex equilibrium configuration. The Magnus force is taken into account.  The average orientation of a vortex line, which corresponds to the direction of a rotation axis of the star, is parallel to the z-axis.  The integration constant in equations (\ref{eq:stst-int1}) is larger in the dashed line than in the solid line.
\label{sinepmt}}
\end{figure}

 Let us move to the case where the integration constant (the "total
energy") in equations  (\ref{eq:stst-int1}) and (\ref{eq:stst-int2}) is smaller than peak values of the sum of pinning and Magnus terms (the "potential energy").  This case corresponds to the motion of a particle which is bound in the potential well.  As the particle exhibits a periodic motion, the vortex line goes back and forth between the two neighboring lattice planes (Fig. \ref{sinepmt}).  The average direction of the vortex line and hence the rotation axis of the star are parallel to the z-axis.  Note that the significant energy is required to deform the vortex from the straight line to the sinusoid-like shape ( see the next sub-section for the calculation of the total energy of a vortex). Hence, it may be the straight vortex line which is realized in nature when the rotation axis is parallel to the z-axis or the major axis of a single crystal lattice.

\subsection{Kink Solutions}
 For simplicity and clarity we neglect the Magnus force here.
Let us consider the case where the integration constant (the "total energy") in equations (\ref{eq:stst-int1}) and (\ref{eq:stst-int2}) is larger than the peak value of the pinning term (the "potential energy"). This case corresponds to the motion of a particle which continues to move in one direction with varying speed. The derivative of $\phi$ and $\psi$ with respect to z is not now zero everywhere along a vortex line.  Hence, we examine here the equilibrium vortex configuration when the average orientation of a vortex line and hence the rotation axis of the star are inclined against the z-axis (or the major axis of a single crystal).

\begin{figure}
\epsscale{0.70}
   \plotone{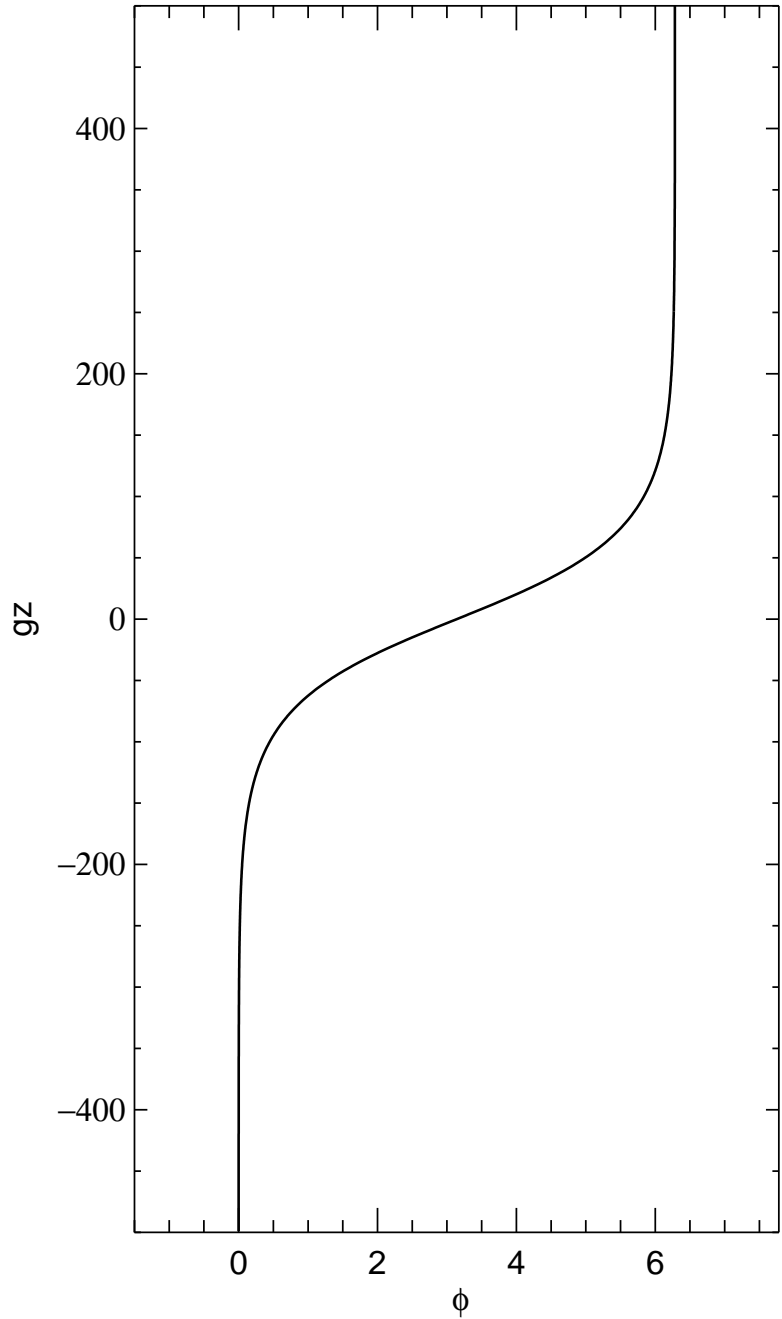}
\caption[f3.eps]{An equilibrium configuration of a vortex line with one kink, derived with no Magnus force.  The vortex moves from one lattice plane ($\phi = 0$) to the adjacent one ($\phi = 2\pi$) by forming a kink.  Except for the kink part, the vortex line is straight and parallel to the z-axis.
\label{kink}}
\end{figure}

We study first the equilibrium solutions when the angle between the rotation axis of a star and the major axis of a crystal is infinitesimally small.  Over an entire interval of a large single crystal the vortex passes from one lattice plane to the adjacent lattice plane.  Hence, we set the boundary conditions such as $\phi = 0$ at $z=-\infty$ and $\phi = 2\pi$ at $z =\infty$.  Then, we can solve equations (\ref{eq:stst-int1}) analytically and derive the equilibrium configurations,
\begin{equation}
 \phi =4\arctan \left[ \exp\left( \frac{z}{l}\right) \right] \label{eq:kink},
\end{equation}
where
\begin{equation}
    l =\sqrt{\frac{\rho \kappa c_K}{4g^2|U_g|}}.\label{eq:kinkl}
\end{equation}
Figure \ref{kink} shows that the vortex can move from one lattice plane to the adjacent lattice plane by bending itself and forming a kink when the vortex tension is finite.  The vortex line is displaced substantially from the pinning sites in the kink part, while in the other part the vortex line is straight and passes through nearly all of the pinning sites along the z-axis.   $l$ given by equation (\ref{eq:kinkl}) expresses a length of the kink.  In the middle of a kink the vortex line is inclined against the z-axis by $2.5^{\circ}$.

Energy is needed to create a kink from a straight vortex line since the length of a vortex line increases and a part of it should overcome the pinning potential.  The self-energy of vortex lines is equal to the kinetic energy of the velocity field induced by them.  The increase of the vortex self-energy due to bending is

\begin{eqnarray}
\Delta E_s=\int_{-\infty}^{\infty} \frac{{\rho}{\kappa}c_K}{2}\left(\frac{\partial {\bf u}}{\partial z} \right)^2 dz\\
=\int_{-\infty}^{\infty} \frac{{\rho}{\kappa}c_K}{4g^2} 
\left[ \left( \frac{\partial \phi }{\partial z} \right)^2 + \left( \frac{\partial \psi }{\partial z} \right)^2 
\right] dz,
\end{eqnarray}
 while the energy necessary to overcome the pinning potential is
 
\begin{equation}
\Delta E_p=\int_{-\infty}^{\infty} 2 U_g (\cos \phi+ \cos \psi)dz-\int_{-\infty}^{\infty} 2 U_g (\cos 0+ \cos 0)dz .
\end{equation}
The length of a kink can be determined from minimizing the sum of $\Delta E_s$ and $\Delta E_p$ \citep{lin91, rud00}.  Using the solution (\ref{eq:kink}) together with $\psi=0$, the total energy necessary to form a kink is calculated as
\begin{equation}
\Delta E_t =\Delta E_s+\Delta E_p=8\left( \frac{|U_g|\rho \kappa c_K}{g^2}\right)^{\frac{1}{2}},\label{eq:et}
\end{equation}
which yields $\sim 4.0 \mbox{ MeV}$ at densities of $\sim 3.4 \times 10^{13} \mbox{ g cm}^{-3}$.

\begin{figure}
\epsscale{0.70}
   \plotone{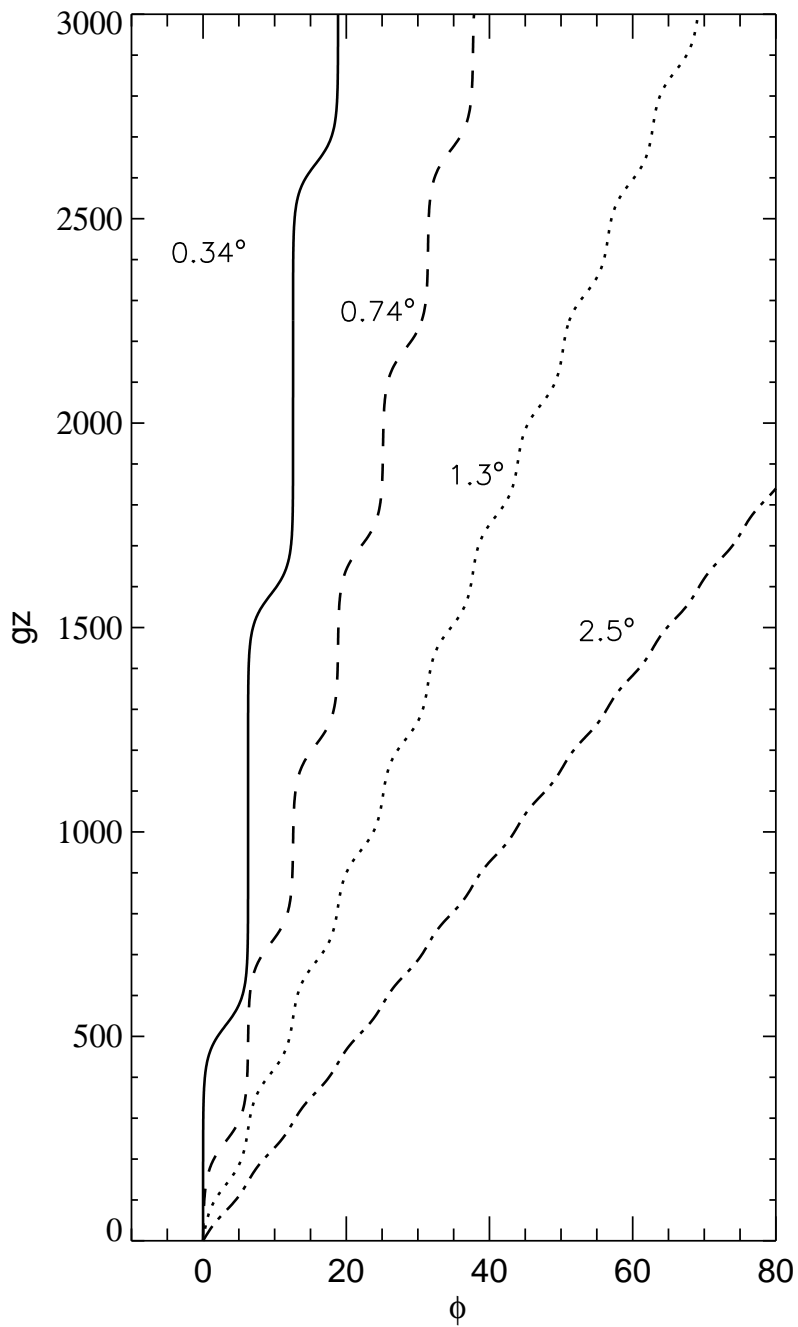}
\caption[f4.eps]{ Vortex equilibrium configurations with kinks for different inclination angles of a rotation axis against the z-axis.  Here the Magnus force is ignored.  The numbers on the curves denote the inclination angle. The average orientation of a vortex line corresponds to the direction of a
rotation axis of the star.  The main lattice planes are located at the
horizontal coordinate of $\phi = 2n\pi$.  As the rotation axis inclines against the z-axis, the pinned straight lengths decrease and kinks become less prominent. \label{stpt}}
\end{figure}

We increase gradually the integration constant in equations (\ref{eq:stst-int1}) and (\ref{eq:stst-int2}) and hence the inclination angle $\theta$ of the rotation axis against the major axis (z-axis) of a single crystal.  The vortex lines now should thread through many lattice planes.  We find that when the inclination angle is small and less than $0.7^{\circ}$, vortex lines in equilibrium configurations move from one lattice plane to the next by forming a clear kink and the number of kinks increases with increasing inclination angle (Fig. \ref{stpt}). The separation $L_z$ in z between the neighboring kinks varies as $L_z \propto \theta ^{-1}$. The property of a kink in Fig. \ref{stpt} is the same as that in Fig. \ref{kink}.  A vortex line can orient its average direction to the direction of the rotation axis by adjusting a number of kinks. Apart from the kink parts, the vortex lines are straight and in parallel to the z-axis, and are firmly pinned to the lattice nuclei.  Even though the extra energy is required to create kinks, the total energy of a vortex line with kinks is much lower than that of the inclined straight vortex line without kinks since except for kink parts the vortex line lies in the bottom region of the pinning potential \citep{lin93}.

As the average orientation of a  vortex line inclines against the major axis,  the separation between the neighboring kinks and hence the portion of a vortex line available for pinning decrease.  As seen in Fig. \ref{stpt}, the straight pinned part of the vortex line becomes shorter than the kink part when the inclination angle is above $\sim 0.7^{\circ}$.  The vertical pinned portion as well as the kink feature becomes less prominent at $\sim 1.3^{\circ}$. When the inclination angle exceeds $\sim 2.5^{\circ}$, the equilibrium configuration of a vortex line becomes almost the straight line.  This consequence arises from the fact that the portion of a vortex line for pinning becomes too short to compensate the energy cost for producing kinks when the inclination angle exceeds a certain value.

When the average orientation of a  vortex line inclines significantly against 
the z-axis, we need to include the z-components for both the force and 
displacement, though neglected here.  We expect that the lattice planes different 
from those considered here may become important for pinning. The three 
dimensional calculations are needed in order to know the equilibrium 
configuration of a vortex line accurately.

\subsection{Effect of the Magnus Force }
We now examine how the kink solutions are modified when we include the Magnus terms in equations (\ref{eq:stst-int1}) and (\ref{eq:stst-int2}).  The pinning and Magnus terms as a function of $\phi$ ($\psi$) are represented by the cosine curve and the inclined straight line, respectively.  Hence, there exists $\phi_c$ ($\psi_c$) at which the integration constant (the "total energy") becomes equal to the sum of the pinning and Magnus terms (the "potential energy").  The integration constant is larger (smaller) than the sum of the pinning and Magnus terms at   $\phi < \phi_c$ ($\psi > \psi_c$), while smaller (larger) at $\phi > \phi_c$ ($\psi < \psi_c$).  This case corresponds to the motion of a particle which moves forwards and then at the potential wall bounces to move backwards.

\begin{figure}
\epsscale{0.70}
   \plotone{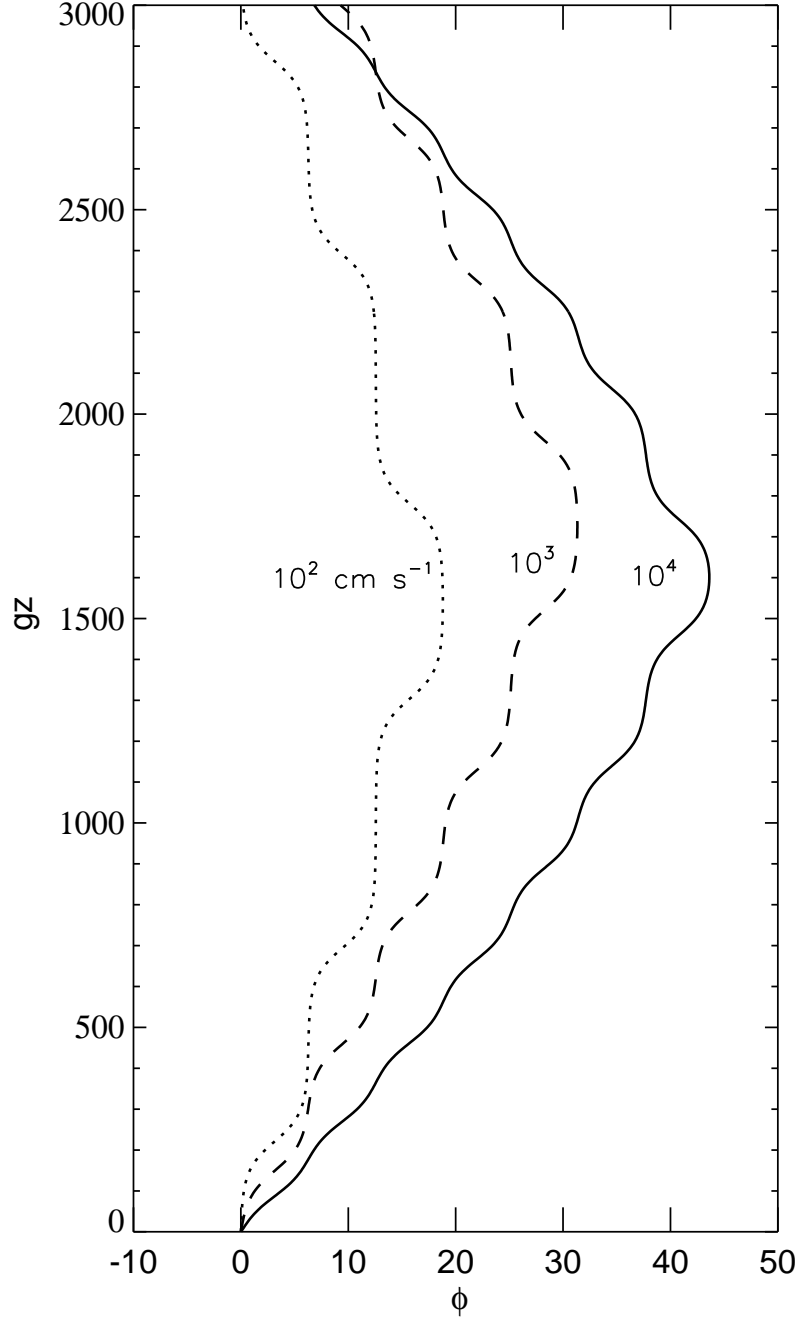}
\caption[f5.eps]{Vortex equilibrium configurations when the Magnus force is included.  The superfluid flows in the y-direction.  The numbers attached to the curves denote the superfluid velocity relative to the lattice nuclei. The boundary conditions adopted are $\phi = 0$ at $z = 0$, and $\partial \phi /g\partial z= 1.4\times 10^{-3}, \, 6.6\times 10^{-3}, \, 2.4\times 10^{-2}$ at $z = 0$ respectively for the cases of $v_s =10^2, 10^3$ and $10^4 \mbox{ cm s}^{-1}$.  The parabolic structure appears on the long scale and its curvature increases with the relative velocity.  As the Magnus force increases, the pinned straight lengths decrease and the kinks become less prominent. \label{kink-parab}}
\end{figure}

 Figure \ref{kink-parab} illustrates the equilibrium configurations of vortex
lines for different superfluid velocities $v_s$ relative to the lattice nuclei. The equilibrium configuration is characterized locally by a kink and globally by a parabola.  As explained above, the kink arises from the balance between the pinning and tension terms. As expected from the analogy with the particle motion, the vortex lines extends first rightwards and then leftwards as z increases.  When we take a long scale average of equation (\ref{eq:eqmot}), the pinning forces, which alternate the sign (the direction) along a vortex line, cancel out whereas the Magnus force, which is constant along a vortex line, is left.  Hence, when combined with the tension term in equation (\ref{eq:eqmot}), the constant Magnus term yields the parabolic configuration. The curvature radius of a vortex line is given approximately by
\begin{equation}
R_c \sim  \left( \frac{\partial^2 u}{\partial z^2} \right) ^{-1}=\frac{c_K}{v_s}.
\end{equation}
As the superfluid velocity relative to the lattice nuclei increases, the
curvature and hence the deformation due to the Magnus force becomes larger
(Fig.  \ref{kink-parab}).  Notice also that the kink feature becomes less prominent 
as the deformation becomes larger due to the same reason as mentioned in the 
previous sub-section. The three dimensional calculations are needed again.

\subsection{Effect of a Density Variation}
 Numerical estimates in this paper are made mostly by adopting the physical parameters at the inner part of an inner crust, where the density is higher and the pinning energy peaks.  The kink properties and pinning strength depend on matter densities through the nature of nuclei and superfluids.  For example, the kink length and creation energy are $l/a \sim 7.3$ and $\Delta E_t \sim 4.0 \mbox{ MeV}$ at densities of $\rho \sim 3.4\times 10^{13} \mbox{ g cm}^{-3}$ in the inner part of an inner crust, while $l/a \sim 39$ and $\Delta E_t \sim 0.089 \mbox{ MeV}$ at densities of $\rho \sim 2.7\times 10^{12} \mbox{ g cm}^{-3}$ in the outer part of an inner crust.  The choice of parameters may affect the results. Note, however, that most of crust superfluids reside in the inner part of an inner crust.  Hence, even if the outer part of the inner crust is included adequately, the change may be small and may not alter the essence of our conclusions.

\subsection{Effect of a Polycrystalline Structure}
The inner crust nuclei are likely to take a polycrystalline structure \citep{jon98}.  The orientation of a crystal lattice may change from one domain to another.  Then, along a vortex line there may lie a large number of single crystals with various orientations against the average direction of a vortex line.  The important physical parameters which vary with the orientation of a crystal lattice are the distance between the neighboring lattice planes for pinning and the distance between the neighboring pinning sites along a vortex line.  It is likely that even when the vortex line is not close to the main axis of a crystal lattice, the equilibrium configuration of a vortex line may also be expressed by a kink solution with the pinning lattice planes different from those containing the main axis of a crystal lattice.  As seen from equations (\ref{eq:kinkl}) and (\ref{eq:et}), the kink length and creation energy are in proportion to the distance between the neighboring lattice planes for pinning and depend weakly on the Fourier coefficient ($U_g$) of a pinning potential.  The kink properties may vary, though not significantly, depending on the orientation of a crystal lattice.  The pinning strength of a vortex line is inversely proportional to the separation of the neighboring pinning sites along a vortex line.  When the vortex line is not close to the major axis of a crystal lattice, the pinning site distance will be longer and hence the pinning of a vortex line will be less strong.  The various orientations of crystal lattice along a vortex line may reduce the pinning strength of a vortex line as a whole compared to the case with a small inclination angle.  We expect,
however,  that pinning, though weaken, is still strong enough to explain the large glitches since the change of the pinning site distance more than an order of magnitude is less likely.  We need to conduct the three dimensional calculations in order to understand the equilibrium configuration and pinning strength accurately for various orientations of crystal lattice as well as for large Magnus forces.

\section{VORTEX OSCILLATIONS}\label{oscillation}
Let us now consider the oscillations excited on and propagate along the vortex line \citep{lin93}.  In order to examine oscillations of small amplitude about equilibrium, we write
\begin{equation}
\phi =\phi_0 +\phi_1\label{eq:phi1}
\end{equation}
and
\begin{equation}
\psi=\psi_0+\psi_1,\label{eq:psi1}
\end{equation}
where the subscripts $0$ and $1$ denote the equilibrium solutions and the
perturbations, respectively.  Using equations (\ref{eq:phi1}) and (\ref{eq:psi1}), we linearize
equations (\ref{eq:eqmot1}) and (\ref{eq:eqmot2}) and obtain
\begin{equation}
4g^2U_g\cos \phi_0 \times \phi_1 -\rho \kappa \frac{\partial \psi_1}{\partial t}+\rho \kappa c_K \frac{\partial^2 \phi_1}{\partial z^2}=0\label{eq:pert1}
\end{equation}
and
\begin{equation}
4g^2U_g\cos \psi_0 \times \psi_1 +\rho \kappa \frac{\partial \phi_1}{\partial t}+\rho \kappa c_K \frac{\partial^2 \psi_1}{\partial z^2}=0\label{eq:pert2}
\end{equation}
We consider plane-wave solutions $\propto \exp(ipz-i\omega t)$ for $\phi_1$ and $\psi_1$, where $\omega$ is the angular frequency and $p$ is the wave number.  Then, we derive the dispersion relation for oscillations,
\begin{equation}
(4 g^2 U_g \cos \phi_0-\rho \kappa c_K p^2)(4 g^2 U_g \cos \psi_0-\rho \kappa c_K p ^2)-(\rho \kappa \omega)^2=0.\label{eq:disp}
\end{equation}
If we adopt such solutions as $\psi_0=- \phi_0$ for equilibrium, from equation (\ref{eq:disp}) we obtain the dispersion relation simplified as
\begin{equation}
\omega = \left| c_K p^2+\frac{4g^2|U_g|}{\rho \kappa}\cos \phi_0 \right| .\label{eq:disp2}
\end{equation}

Next, we examine the property of oscillations represented by the dispersion relation (\ref{eq:disp2}).  The first term in the dispersion relation (\ref{eq:disp2}) can be derived when in equations (\ref{eq:pert1}) and (\ref{eq:pert2}) we include only the second and third terms ( the Magnus and tension terms) ignoring the first term (the pinning term).  Hence, the first term in the dispersion relation (\ref{eq:disp2}) expresses the Kelvin wave, which is circularly polarized and propagates helically along a vortex line \citep{son87}.  The second term in the dispersion relation (\ref{eq:disp2}) can be derived when in equations (\ref{eq:pert1}) and (\ref{eq:pert2}) we include only the first and second terms (the pinning and Magnus terms) neglecting the third term (the tension term).  We find that the second term in the dispersion relation expresses the circular motion of a vortex line as a whole around the equilibrium position.  In general the dispersion relation (\ref{eq:disp2}) represents the combined oscillations of Kelvin and rotational modes.
       
\begin{figure}
\epsscale{0.70}
   \plotone{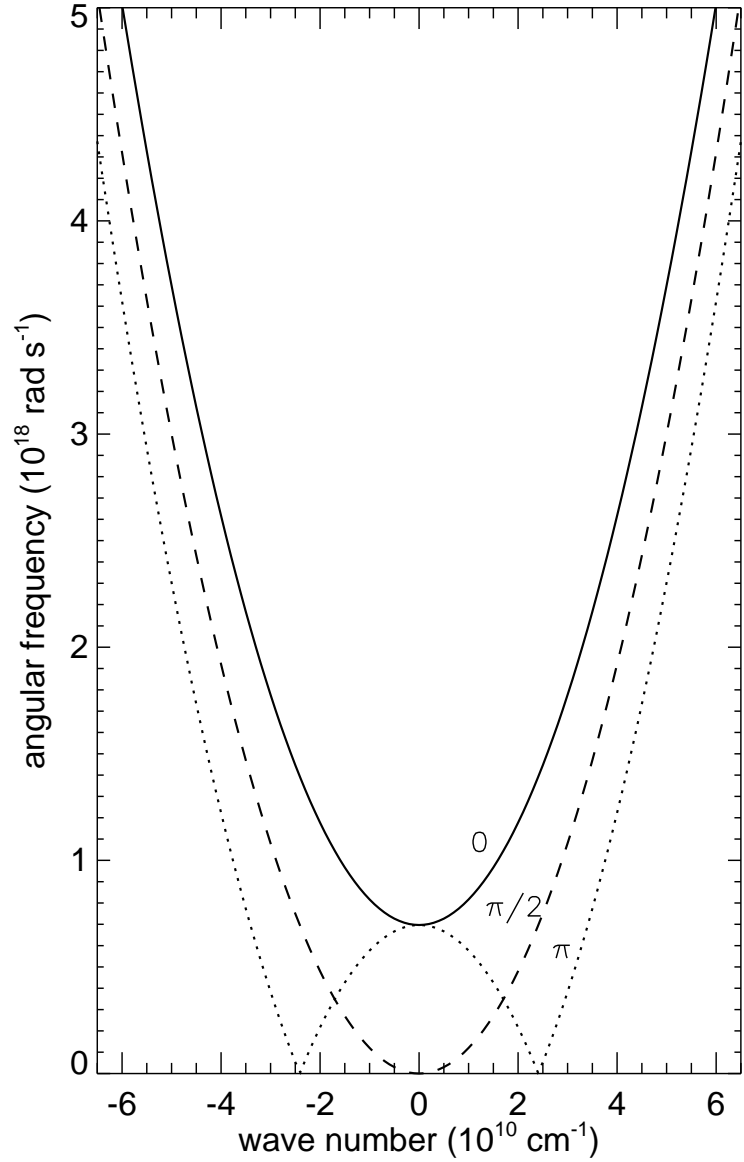}
\caption[f6.eps]{Dispersion relations for oscillations allowed on a vortex line. The numbers by the curves denote a displacement of the vortex segment, $\phi_0$, from the z-axis in equilibrium configuration.  The oscillations in the pinned segments are possible only above a minimum frequency. \label{disp}}
\end{figure}

 Figure \ref{disp} illustrates the dispersion relation given by equation
(\ref{eq:disp2}) for different $\phi_0$.  The angular frequency of allowed oscillations is plotted as a function of wave number.  At smaller wave numbers the angular frequency of oscillations is determined mainly by the rotational mode, while at larger wave numbers mainly by the Kelvin mode.  We should note that there exists the minimum frequency for the oscillations especially when  $\phi_0$ is close to $2n\pi $. The minimum frequency for $\phi_0 = 2n\pi$ is
\begin{equation}
\omega_{min}=\frac{4g^2|U_g|}{\rho \kappa}\label{eq:wmin}
\end{equation}
which yields $7 \times 10^{17} \mbox{ rad s}^{-1}$ at densities of $\sim 3.4 \times 10^{13} \mbox{ g cm}^{-3}$ and $1 \times 10^{16} \mbox{ rad s}^{-1}$ at densities of $\sim 2.7 \times 10^{12} \mbox{ g cm}^{-3}$. $\phi_0 \sim 2n\pi$ in the pinned segments, where the vortex line lies deep in the pinning potential well.  On the other hand, at the kink segments where $\phi_0 \sim (2n+1)\pi$, the minimum frequency is zero.  These results lead to the important conclusion that the vortex oscillations with frequencies lower than the minimum frequency, $\omega \lsim 7 \times 10^{17} \mbox{rad s}^{-1}$, can not be excited on and can not propagate in the pinned part of a vortex line, whereas oscillations with frequencies higher than the minimum can propagate along a vortex line, varying the wave length.  The vortex oscillations with lower frequencies are allowed in the kink part of a vortex line.  We should note, however, that the dispersion relation (\ref{eq:disp2}) in the kink part is less reliable for lower frequencies since the wave length exceeds the kink length and hence the local perturbation analysis is not adequate.  We need to conduct the non-local analyses for the accurate dispersion relation in low frequencies.

In a polycrystalline structure the vortex line passes through a number of single crystals with various orientations. The local perturbation analysis can be applied to each single crystal to examine the vortex oscillations. Equations (\ref{eq:disp2}) and (\ref{eq:wmin}) may also be derived for each single crystal, although the physical parameters in one domain differ from those in another.  We can expect the physical parameters to change rather sharply at the boundaries between the single crystals. If the frequencies are sufficiently high above the minimum frequencies, the oscillations will propagate along a vortex line passing through many single crystals. If the frequencies are around the minimum frequencies, the oscillations are reflected at the boundaries and in some cases will be trapped within one crystal or some. If the frequencies are lower than the minimum frequencies, there may be no oscillations which propagate over a number of single crystals.

\section{VORTEX PINNING}\label{pinning}
It has been considered that pulsar glitches are a consequence of the large scale unpinning of superfluid neutron vortices from the lattice nuclei in the inner crust of neutron stars.  Recently, however, Jones (1997, 1998, 1999) argues that vortex interaction with a polycrystalline structure does not provide pinning strong enough to explain the large glitches observed in the Vela pulsar.  Let us examine this conclusion in the light of our new results mentioned in the previous sections.

Jones describes the reason for the conclusion that the structural regularity of a defect-free single crystal allows displacement of a vortex, under a Magnus force, to a continuous sequence of new positions with no more than extremely small changes in energy.  Furthermore, Jones
considers the motion of kinks propagating along a vortex line with a velocity of $v_K$.  For simplicity we consider the vortex lines which have kinks bent in the y-direction and extend approximately in the z-direction. The kink propagation along a vortex line yields a displacement  and hence motion of the vortex line in the y-direction.  If the kink propagation velocity is
\begin{equation}
     v_K = \frac{L_z}{a} v_s,\label{eq:vk}
\end{equation}
the time average of a vortex velocity in the y-direction is equal to the superfluid velocity \citep{jon97, jon98}. In effect, vortex lines move in the y-direction together with the superfluid, almost not suffering from pinning.  These energy and velocity arguments lead Jones to the above conclusion that pinning would be extremely weak and insufficient for the large glitches.

We point out several difficulties included in these arguments. Kinks should be created and supplied around the edge of a single crystal at a rate of one kink every $a/v_s$ seconds in order for a vortex line to move steadily together with superfluid.  The kink supply rate required just before the glitches in the Vela pulsar is  $\sim 2\times 10^{15} \mbox{ s}^{-1}$.  This rate
is much lower than the minimum frequency (equation (\ref{eq:wmin})) of oscillations which can be excited in the pinned part of a vortex line.  Hence, the
oscillations of vortex lines cannot be a direct cause for the creation and supply of kinks.  Moreover, energy of $\sim 4.0 \mbox{ MeV}$ (equation (\ref{eq:et})), which is substantially larger than the thermal energy, is needed locally to form a kink.  Energy source for kink formation is also pointed out as a difficulty
included in the kink propagation model.  We note that the kink solution (\ref{eq:kink}) can also be expressed as a sum of fourier components of different wave numbers.  As seen from the dispersion relation (\ref{eq:disp2}), the phase velocity of a vortex wave depends on wave number.  The vortex waves with different wave numbers propagate with different phase velocities.    Hence, even if a kink is formed around the end of a single crystal, the kink feature will be smeared out during propagation. Furthermore, our perturbation analyses show that the vortex configurations with static kink structures are stable.  These facts indicate that the kink motion following equation (\ref{eq:vk}) is less likely.  If the kinks do not move along a vortex line with the velocity given by equation (\ref{eq:vk}), the conclusion by Jones is not guaranteed.

 We show in the previous sections that the vortex equilibrium configuration is in general composed of the static kink and straight segments.  A vortex line in equilibrium lies deep in the pinning potential well and is strongly pinned to the lattice nuclei in its most part except for the kink part.  From the perturbation analyses we find no unstable mode that grows with time for the pinned equilibrium configurations both with and without kinks.  The pinned equilibrium configurations, though oscillations can be excited, may be stable.  Hence, the static kink is a stable structure.  Our studies show that strong pinning of vortex lines is likely to occur especially when the average direction of a vortex line is parallel or approximately parallel to the major axis of a crystal lattice and the
Magnus force is not so strong. The orientation of a crystal lattice varies from one domain to another in a polycrystalline structure.  The pinning strength decreases in the domain where the major axis of a crystal lattice is not close to the vortex line.  However, as long as the equilibrium vortex configurations are represented by the kink solutions, pinning of a vortex line in a polycrystalline structure would be still strong enough to explain the large glitches (see the subsection 3.7).

\section{CONCLUDING REMARKS}\label{conclusion}
We have shown that the equilibrium configuration of a vortex line consists of straight and kink parts when the rotation axis of the star is approximately aligned with the major axis of a crystal lattice and the Magnus force is weak.  The straight part is much longer than the kink part. Furthermore, the straight parts pass through nearly all of nuclei along the major axis, whereas the kink parts lie mostly in space between the lattice planes. The vortex line is strongly pinned to the lattice nuclei in the straight part, while not in the kink part.  We have found from the perturbation analyses that there exist two modes, rotational and helical (Kelvin), for oscillations excited on a vortex line. The vortex oscillations are possible only above a minimum frequency \citep{lin93} and the phase velocity of oscillations varies with wave number. We find no unstable mode that grows with time. These facts lead to the conclusion that the vortex configuration with kinks may be stable and yield the strongly pinned state especially when the vortex line is close to the main axis of a crystal lattice.

 Jones (1997, 1998) argues that pinning is not strong enough to explain the large glitches adopting a kink motion in a polycrystalline structure. Our studies show that the static kink is a stable structure and the kink motion is less likely. If the kinks do not move along a vortex line, the conclusion made by Jones is not guaranteed.  Nuclei in the crust may be in a polycrystalline structure. The orientation of a crystal lattice varies from one domain to another. The pinning strength decreases in the domain where the major axis of a crystal lattice is not close to the vortex line. However, as long as the equilibrium vortex configurations are represented by the kink solutions, pinning of a vortex line in a
polycrystalline structure would be still strong enough to explain the large glitches.

We have shown that the Magnus force makes the vortex line parabolic on global scale and causes it to incline against the major axis of a crystal lattice. As the Magnus force increases and the rotation axis inclines against the major axis of a crystal lattice, we need to include the major axis components for force and displacement, though neglected in the present study. We expect that the lattice planes different from those considered here may become important for pinning.

The pinning sites are discrete although we have assumed the uniform pinning potential along the major axis of a crystal lattice. Link, Epstein and Baym (1993) find that there appears the band structure in the energy spectrum for excitations on a vortex line when the discrete nature is included in the pinning site distribution. It is also necessary to take account of the discrete nature of the pinning sites when we consider the vortex equilibrium configurations with the scale length comparable to the lattice constant.

The above arguments suggest that the three dimensional calculations adopting the discrete pinning sites should be conducted in near future in order to understand the equilibrium vortex configurations and vortex oscillations and judge the relevance of vortex pinning for the pulsar glitches correctly.

\acknowledgments
We thank M. Ruderman, R. Epstein and B. Link for many useful discussions. We also thank S. Nakamura for his help in calculations. This research was supported in part by the Grant-in Aid for Scientific Research (C) (10640234, 12640229, 12640302).

\clearpage

\clearpage

\clearpage

\clearpage

\clearpage

\clearpage

\end{document}